\documentclass[showpacs,superscriptaddress,twocolumn,prl]{revtex4}
\usepackage{amsmath,graphicx}

\begin{document}

\title {Emission and Absorption quantum noise measurement with an on-chip resonant circuit}
\author{J. Basset}
\affiliation{Laboratoire de Physique des Solides, Univ. Paris-Sud, CNRS, UMR 8502, F-91405 Orsay Cedex, France.}
\author{H. Bouchiat}
\affiliation{Laboratoire de Physique des Solides, Univ. Paris-Sud, CNRS, UMR 8502, F-91405 Orsay Cedex, France.}
\author{R. Deblock}
\affiliation{Laboratoire de Physique des Solides, Univ. Paris-Sud, CNRS, UMR 8502, F-91405 Orsay Cedex, France.}

\begin{abstract}
Using a quantum detector, a superconductor-insulator-superconductor junction, we probe separately the emission and absorption noise in the quantum regime of a superconducting resonant circuit at equilibrium. At low temperature the resonant circuit exhibits only absorption noise related to zero point fluctuations whereas at higher temperature emission noise is also present. By coupling a Josephson junction, biased above the superconducting gap, to the same resonant circuit, we directly measure the noise power of quasiparticles tunneling through the junction at two resonance frequencies. It exhibits a strong frequency dependence, consistent with theoretical predictions.
\end{abstract}

\maketitle 

Whereas electrical noise has been extensively studied at low frequency in various systems \cite{blanter00}, going from macroscopic to mesoscopic scales, and is now relatively well understood, investigation of high frequency noise is much more recent. Of particular interest is  the frequency range of the order of or higher than the applied voltage $V$ or temperature $T$ characteristic energy scales.  Current fluctuations in this quantum regime  acquire a frequency dependence with signatures of the relevant energy scales $k_B T$ and $eV$. Thus current fluctuations has been found to increase linearly with frequency above $k_BT/h$ \cite{koch81,yurke88}.  Similarly  the  excess noise, \textit{i.e.} the difference between the noise at a given bias and the noise at equilibrium, measured in the limit $eV \gg k_BT$, has been found to decrease linearly with frequency and go to zero at frequency $eV/h$ both in diffusive wires \cite{schoelkopf97}, tunnel junction \cite{gabelli08} and quantum point contacts \cite{eva07}. In the quantum regime noise can be described in terms of exchange of photons of energy $h \nu$ between the source and the noise detector. Depending on whether photons are emitted or absorbed by the source one measures emission noise (corresponding to negative frequencies) and absorption noise (corresponding to positive frequencies) \cite{clerk10}. This  difference between emission and absorption processes is well known in quantum optics but difficult to observe in electronic devices since most classical amplifiers exchange energy with the measured device and allow only the detection of a combination of emission and absorption noise \cite{lesovik97}. To measure non-symetrized noise, \textit{i.e.} distinguish between emission and absorption one can use a quantum detector \cite{clerk10,aguado00}. Different realizations of such a detection scheme have been implemented using \textit{e.g.} quantum bits \cite{schoelkopf02}, quantum dots \cite{onac06,gustavsson07}, superconductor-insulator-superconductor (SIS) tunnel junction \cite{deblock03,billangeon06} or superconducting resonator \cite{xue09}. Due to the difficulty to extract the equilibrium noise contribution, this type of measurement has only been done so far for the excess noise.

In this Letter we have embedded the tested device, a Josephson junction, in an on-chip superconducting resonant circuit, as in Ref. \cite{xue09,holst94}, and incorporated a quantum detector, a SIS junction, in the same circuit. We thus detect, in the quantum regime $h \nu \gg k_B T$, the emission and absorption noise of the resonator at \textit{equilibrium} and the excess noise of the probed device at the resonance frequencies. At the frequencies probed in the experiment, at low temperature the resonator does not emit noise whereas it shows absorption noise related to its zero point fluctuations. This technique also allows a direct extraction of the excess noise power of quasiparticles tunneling through a Josephson junction at 28.4 and 80.2 GHz, the resonance frequencies of the resonator.
\begin{figure}[tb]
  \begin{center}
		\includegraphics[width=8cm]{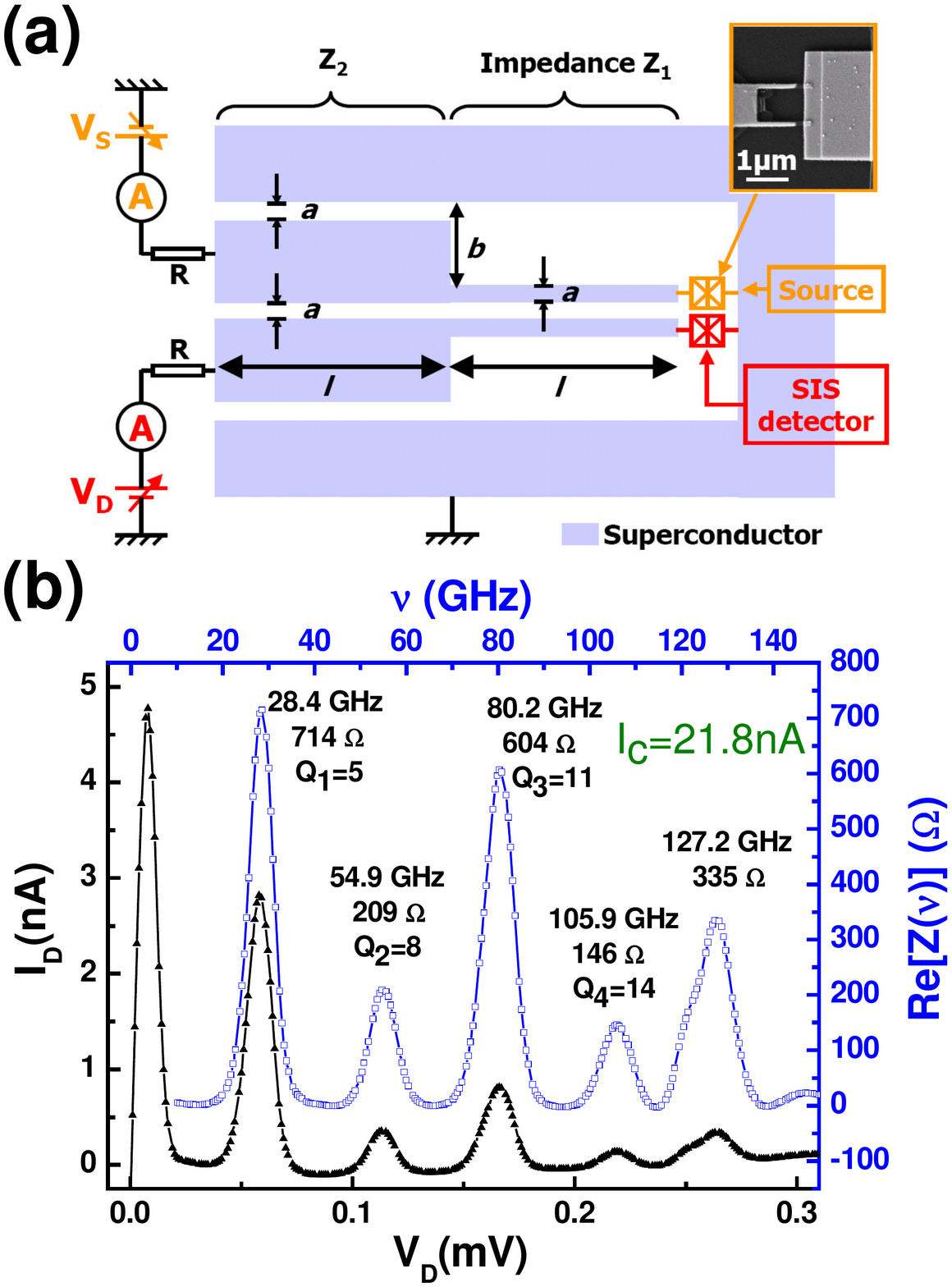}
	\end{center}
  \caption{a- Sketch of the sample, with $a=5\mu$m, $b=100\mu$m and $l=1$mm. The two transmission lines are terminated by on-chip Pt resistors ($R=825 \Omega$). b- Lower curve~: $I(V)$ of the detector in the subgap region with $I_C$ maximized by adjusting the magnetic flux. Upper curve~: the real part of the impedance seen by the detector, extracted using Eq. \ref{Ijj}, exhibits resonances with quality factor $\approx$ 10.}
  \label{fig1}
\end{figure}

The probed device consists of two coupled coplanar transmission lines. Each transmission line is connected to the ground plane via a superconducting tunnel junction of size $240\times150$ nm$^2$ and consists of two sections of same length $l$ but different widths, thus different characteristic impedance $Z_1 \approx 110 \Omega$ and $Z_2 \approx 25 \Omega$(Fig. \ref{fig1}a). Due to the impedance mismatch the transmission line acts as a quarter wavelength resonator, with resonances at frequency $\nu_n=nv/4l=n\nu_1$, with $v$ the propagation velocity and $n$ an odd integer \cite{holst94}. The two transmission lines are close to one another to provide a good coupling at  resonance and are terminated by on-chip Pt resistors. The junctions have a SQUID geometry with different areas to tune separately their critical currents with a magnetic flux. The junctions and the resonator are fabricated in aluminum (superconducting gap $\Delta =260 \mu$eV) on an oxidized silicon wafer. The system is thermally anchored to the cold finger of a dilution refrigerator of base temperature 20 mK and measured through filtered lines with a standard low frequency lock-in amplifier technique. 

Hereafter we call one junction the detector and the other the source. Coupling the detector to the source with a resonant circuit has two main effects on the detector. First, when the source is not biased, the $I(V)$ characteristic of the detector is modified due to the resonator. This allows extracting the noise properties of the resonant circuit at \textit{equilibrium}. Second, when the source is biased the detector is also sensitive to the source noise at the resonance frequencies of the resonator.
\begin{figure}[tb]
	\begin{center}
		\includegraphics[width=8cm]{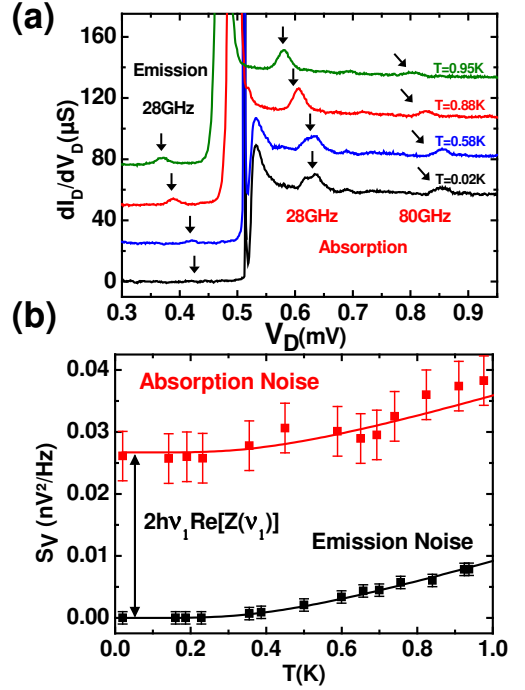}
	\end{center}
	\caption{a- $dI/dV_D$ of the detector at different temperatures with $I_C$ minimized by adjusting the magnetic flux. The curves are shifted vertically for clarity. The peaks corresponding to the detection of emission or absorption noise are denoted by arrows. b- Dependence versus T of the power of voltage noise at $\nu_1=28.4$GHz in emission and in absorption. The solid line corresponds to the theoretical prediction (Eq. \ref{SV}). Only absorption noise is detected below 0.4 K.}
	\label{fig2}
\end{figure}

First we neglect the effect of the source junction, on which no bias voltage is applied. The $I(V)$ characteristic of a small Josephson junction depends on the impedance of its electromagnetic environment \cite{ingold92}. For the case of a superconducting transmission line resonator \cite{holst94}, resonances appear in the subgap region $V_D<2\Delta/e$ due to the excitation of the resonator modes by the AC Josephson effect \cite{tinkham96,barone82}. These resonances are related to the real part of the impedance $Z(\nu)$ seen by the junction~:
\begin{equation}
	I(V)=Re[Z(2eV/h)] \, I_C^2/2V 
	\label{Ijj}
\end{equation}
with $I_C=\pi \Delta/(2eR_T)$ \cite{tinkham96} the critical current and $R_T=18.7k\Omega$ the normal state resistance of the junction. Eq. \ref{Ijj} can be derived as the effect of the electromagnetic environment on the tunneling of Cooper pairs through the Josephson junction \cite{ingold92} or by writing that the DC power provided by the voltage source to the Josephson junction is equal to the AC power dissipated by the resistive part of the environment $Re[Z(\nu)]$ due to the AC Josephson effect at the Josephson frequency $\nu=2eV/h$. Fig. \ref{fig1}b shows the $I(V)$ characteristic of the junction in the subgap region for $I_C$ maximized with magnetic flux. Using Eq. \ref{Ijj} the subgap resonances allow to extract the real part of the impedance seen by the junction. It exhibits peaks at frequencies $\nu_{1,2,3}=28.4,54.9$ and $80.2$ GHz. With a length $l=1$mm the first resonance was expected at 30 GHz. We attribute the difference with the measured resonance frequency to the capacitances of the junctions, of the order of 3.5fF, which shift the resonance. The relatively low quality factors $Q_n$, typically 10, are related to the unavoidable direct connection of the biasing circuit to the transmission line. The fact that we see resonances at frequencies $\nu_n=nv/4l$, with $n$ not only odd but also even is attributed to the rather small ratio $Z_1/Z_2 < 10$ of the impedances of the transmission lines. 
\begin{figure}[tb]
	\begin{center}
		\includegraphics[width=8cm]{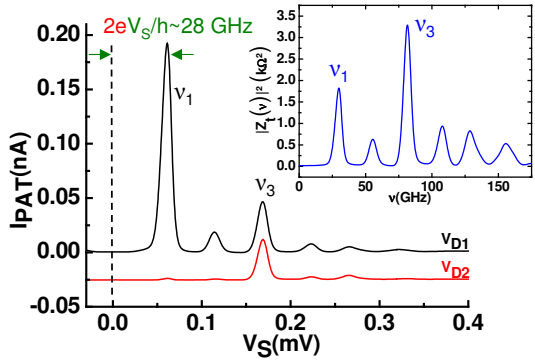}
	\end{center}
	\caption{Calibration using AC josephson effect : the PAT current through the detector versus $V_S$. The curves are shifted vertically for clarity. $V_D$ selects the detected noise frequencies $\nu$ of interest : the upper curve is taken at $V_{D1}=450 \mu$V, corresponding to $\nu \geq (2 \Delta-eV_{D1})/h = 17$ GHz, the lower curve, at $V_{D2}=300 \mu$V, corresponds to $\nu \geq 53$ GHz. Inset : Frequency dependence of the coupling $|Z_t(\nu)|^2$ deduced from the curve taken at $V_{D1}$.}
	\label{fig3}
\end{figure}

The resonant circuit coupled to the junction also leads to a photo-assisted tunneling (PAT) quasiparticle current through the detector. To probe this effect the magnetic flux is adjusted to minimize the critical current of the detector. The dependence of quasiparticles tunneling in SIS junction versus irradiation with microwave photons has been widely used for making mixers \cite{tucker85}. More recently SIS junctions have been used as quantum detectors of noise in mesoscopic physics \cite{deblock03,billangeon06}. The PAT current through the detector as a function of bias voltage $V_D$ and non-symmetrized spectral density of voltage noise $S_V(\nu)$ across the junction \cite{ingold92,aguado00} reads for $k_B T \ll eV_D$ and small voltage noise $S_V(\nu)$ :  
\begin{eqnarray}
	\nonumber I_{PAT}(V_D) &=&\int^{\infty}_{0} d\nu \left( \frac{e}{h\nu} \right)^2 S_V(-\nu)I_{QP,0}(V_D+\frac{h\nu}{e})\\
	\nonumber &+&\int^{eV_D/h}_{0} d\nu \left( \frac{e}{h\nu} \right)^2 S_V(\nu)I_{QP,0}(V_D-\frac{h\nu}{e})\\
 	&-&\int^{+\infty}_{-\infty} d\nu \left( \frac{e}{h\nu} \right)^2 S_V(\nu)I_{QP,0}(V_D)
	\label{IPAT}	
\end{eqnarray}
with $I_{QP,0}(V_D)$ the $I(V)$ characteristic of the detector without electromagnetic environment. The first term of eq. \ref{IPAT} is related to the emission noise, the second to the absorption noise and the third renormalizes the elastic current. When $|V_D|<2 \Delta/e$, due to the superconducting density of states, $I_{QP,0}(V_D)=0$ so that only emission noise is detected for frequencies higher than $(2 \Delta-e V_D)/h$. For $|V_D| > 2 \Delta/e$  the detector is mainly sensitive to absorption noise. To probe more accurately the small effect of the resonant circuit on the detector we modulate at 13.3 Hz $V_D$ and detect by a lock-in technique $dI/dV_D$ at different temperatures (Fig. \ref{fig2}a). At low temperature, on top of the expected $dI/dV_D$ curve of the detector, we see peaks (Fig. \ref{fig2}a) at $eV_D=2 \Delta + h\nu_n$ with $\nu_n$ the resonance frequencies of the circuit coupled to the detector. No such peaks are detected for $e V_D < 2 \Delta$ but, at higher temperature between 200 mK and 1K, a peak at $e V_D=2 \Delta -h \nu_1$ appears and grows with temperature. The position in $V_D$ of the peaks changes due to the temperature dependence of the superconducting gap. Higher temperature were not considered due to the strong temperature dependence of the SIS detector for $T > 1$K. These peaks in the $dI/dV$ characteristics of the detector are attributed to its sensitivity to the voltage fluctuations of the resonant circuit~: 
\begin{equation}
	S_{V}(\nu,T)=2Re[Z(\nu)]h\nu/(1-\exp{(- h\nu/k_B T)})
	\label{SV}
\end{equation}
Eq. \ref{SV} describes the crossover between thermal noise at low frequency and quantum noise related to the zero point fluctuations of the electromagnetic field. The resonances of $Re[Z(\nu)]$ at $\nu_n$ thus cause noise peaks at frequencies $+\nu_n$ (absorption, with $\nu_n >0$) and $-\nu_n$ (emission). At low $T$ only peaks in absorption are predicted whereas when $T$ increases peaks in emission should appear. For $e V_D < 2\Delta$ only the first term in eq. \ref{IPAT} is non-zero. For a peaked emission noise at $-\nu_n$, approximating the integral by a sum and noting $\delta \nu_n=1.06 \nu_n/Q_n$, related to the width of the resonance at frequency $\nu_n$ \cite{delta_nu_n} extracted from Fig. \ref{fig1}b, yields~:
\begin{equation}
	I_{PAT}(V_D)=\sum_n \frac{e^2 S_V(-\nu_n) \delta\nu_n}{(h\nu_n)^2} I_{QP,0}(V_D+\frac{h\nu_n}{e})
	\label{IpatEm}
\end{equation}
Only the absorption term in eq. \ref{IPAT} leads to peaks in $dI/dV$ for $V_D > 2\Delta/e$~: 
\begin{equation}
	I_{PAT}(V_D)=\sum_n \frac{e^2 S_V(\nu_n) \delta\nu_n}{(h\nu_n)^2} I_{QP,0}(V_D-\frac{h\nu_n}{e})
	\label{IpatAb}
\end{equation}
From eq. \ref{IpatEm} and \ref{IpatAb} we extract the emission and absorption voltage fluctuations of the resonant circuit at $\nu_1=28.4$ GHz at different $T$ (Fig. \ref{fig2}b). To do so we integrate the corresponding peak in $dI/dV_D$, at $V_D=(2 \Delta - h \nu_1)/e$ for emission and $V_D=(2 \Delta + h \nu_1)/e$ for absorption, to obtain the value of $I_{PAT}$. $\delta \nu_1=6.66$ GHz is extracted from Fig. \ref{fig1}b and the current $I_{QP,0}(V)$ is measured from the $I(V)$ of the detector at temperature $T$ \cite{IV}. The same treatment is done for the absorption noise at $\nu_3=80.2$Ghz and leads to $S_V(\nu_3)=0.062 \pm 0.005$nV$^2/$Hz between 20 and 950 mK, consistent with the expected value $0.064$nV$^2/$Hz. The temperature dependence of voltage fluctuations agrees quantitatively with theoretical predictions (eq. \ref{SV}). Deep in the quantum regime ($h \nu_1 \gg k_B T$) the voltage fluctuations at \textit{equilibrium} are dominated by the zero point fluctuations of the electromagnetic field and do not exhibit any emission noise. For $h \nu_1 \geq k_B T$ the crossover to thermal noise is visible.
\begin{figure}[tb]
	\begin{center}
		\includegraphics[width=8cm]{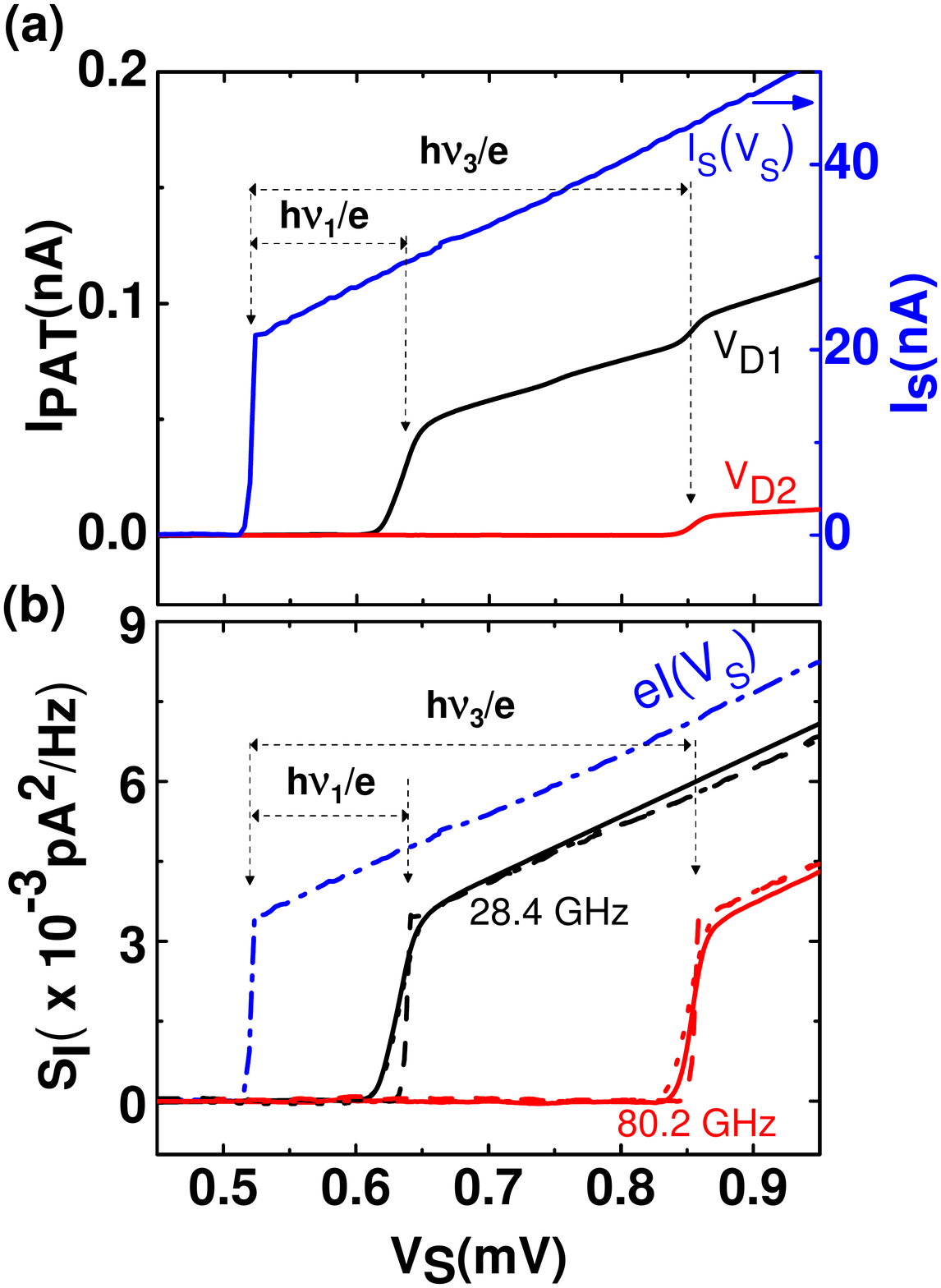}
	\end{center}
	\caption{a- PAT current through the detector versus $V_S$, with source biased on the quasiparticle branch. The curves are taken at $V_{D1}$ and $V_{D2}$. The $I(V)$ of the source is also shown. b- Extracted noise power in emission at $\nu_1=28.4$ GHz and at $\nu_3=80.2$ GHz. For comparison the expected noise power is plotted (dashed curves) along with the noise average over the bandwidth $\delta\nu_n$ of detection (dotted lines). The agreement is within 5\%. The expected zero frequency emission noise power is plotted (dashed-dotted line).}
	\label{fig4}
\end{figure}

We have neglected so far the source junction. When it is biased it can emit noise, which couples to the detector \textit{via} the resonant circuit. To probe the emission of the source the detector is biased at $V_D < 2 \Delta/e$ and the source bias voltage $V_S$ is modulated so as to detect by a lockin technique only the source contribution to the PAT current, \textit{i.e.} $\partial I_{PAT}/\partial V_S$. This quantity can then be numerically integrated to measure $I_{PAT}(V_S)$. To characterize the coupling between source and detector we use the AC Josephson effect of the source for calibration. On Fig. \ref{fig3} the PAT current through the detector versus $V_S$ is shown at two detector voltages $V_{D1}= 450 \mu$V and $V_{D2}= 300 \mu$V. In this regime where the detector is irradiated by the Josephson current at frequency $\nu=2eV_S/h$ the PAT current reads \cite{tien63}~: 
\begin{equation}
I_{PAT}(V_D)=e^2|Z_t(\nu)|^2 I_C^2 I_{QP}(V_D+h\nu/e)/4(h\nu)^2
\label{IjAC}
\end{equation}
with $I_C$ the critical current of the source junction, $Z_t(\nu)$ the transimpedance measuring the coupling between the source and the detector and $I_{QP}(V_D)$ the $IV$ characteristic of the detector. Using eq. \ref{IjAC} we can extract from the curve measured at $V_{D1}$ the value of the coupling $|Z_t(\nu)|^2$ (Fig. \ref{fig3}). It exhibits resonances at the same frequencies as the resonator. This scheme allows a rather strong coupling, proportional to the quality factor of the resonance of $|Z_t(\nu)|^2$, for a finite number of frequencies. This contrasts with previous experiments using a capacitive coupling between source and detector \cite{deblock03,billangeon06}, leading to a relatively small coupling over a wide range of frequencies.

When $e V_S \geq 2 \Delta$ the noise due to the tunneling of quasiparticles can be probed. The PAT current in this regime is shown on Fig. \ref{fig4} for two values of detector bias $V_{D1}$ and $V_{D2}$. We use eq. \ref{IpatEm}, with $S_V(\nu)=|Z_t(\nu)|^2 S_{I_{QP}}(\nu,V_S)$ and $\delta \nu_n$ the width of the resonances of $|Z_t(\nu)|^2$, to extract quantitatively the noise spectrum from Fig. \ref{fig4}. When $V_D < 2 \Delta /e$, only the frequencies higher than $(2 \Delta -e V_D)/h$ need to be considered in the sum. Consequently for $V_D=V_{D1}$ the detector is mainly sensitive to the noise at frequencies $\nu_1$ and $\nu_3$, whereas for $V_D=V_{D2}$ only the noise at $\nu_3$ is detected. The noise at $\nu_1$ is thus extracted from $I_{PAT}(V_{D1})-I_{PAT}(V_{D2})$. One then obtain the spectral density of quasiparticle noise in emission at $\nu_1=28.4$GHz and $\nu_3=80.2$GHz (Fig. \ref{fig4}). We compare these results to the theoretical prediction~:
\begin{equation*}
	S_{I_{QP}}(\nu,V_S)=e \left[\frac{I_{QP}(h\nu/e+V_S)}{1-e^{-\beta(h\nu+eV_S)}}+\frac{ I_{QP}(h\nu/e-V_S)}{1-e^{-\beta(h\nu-eV_S)}} \right]
\end{equation*}
and to the noise integrated over the detection bandwidth $\delta\nu_n$. The agreement is within 5\% in amplitude with this last quantity and the frequency dependence is well reproduced. This is a direct quantitative measurement in the quantum regime $h \nu \gg k_B T$ of the quantum noise associated with the quasiparticles tunneling.
   
In conclusion we have measured the emission and absorption noise of a resonant circuit at \textit{equilibrium} by coupling it to a quantum detector, a SIS junction. At low temperature the circuit exhibits only absorption noise related to the zero point fluctuations of the electromagnetic field. At higher temperature emission noise is also present. The design of the resonant circuit allows to couple another device to the detector and to measure quantitatively its noise at the resonances of the resonant circuit with an accuracy proportional to their quality factors. For a Josephson junction biased above the superconducting gap it was thus possible to probe quantitatively the quasiparticle current noise at 28.4GHz and 80.2GHz and in particular its strong frequency dependence. This technique can be used to probe quantum noise of relatively resistive mesoscopic devices at high frequency.

We acknowledge M. Aprili, B. Reulet, J. Gabelli, I. Safi, P. Simon, C. Ojeda-Aristizabal, M. Ferrier and S. Gu{\'e}ron for fruitful discussions. This work has benefited from financial support of ANR (contract ANR-09-BLAN-0199-01) and C'Nano IdF (project HYNANO).


\begin{thebibliography}{99}
\bibitem{delta_nu_n}
$\delta \nu_n$ multiplied by the amplitude of the resonance peak is equal to the area of the peak.
\end{thebibliography}
\end{document}